\documentclass[final, 3p, times]{elsarticle}
\usepackage[T1]{fontenc}
\usepackage{amsmath,amssymb,bm,booktabs,tabularx,array,longtable,graphicx,listings,float}
\usepackage{microtype}
\usepackage{xurl}
\usepackage[hidelinks,unicode]{hyperref}
\newcommand{\dd}{\mathrm{d}}
\newcommand{\code}[1]{\texttt{\detokenize{#1}}}

\newcommand{\doublefigurewidth}{0.96\textwidth}
\makeatletter
\def\ps@pprintTitle{\let\@oddhead\@empty\let\@evenhead\@empty\def\@oddfoot{\hfil\thepage\hfil}\let\@evenfoot\@oddfoot}
\makeatother
\begin{document}
\begin{frontmatter}
  \title{VEQDB: A Compact and Reconstructible Multi-Device Tokamak Equilibrium Database}
  \author[veloalpha]{Huasheng Xie\corref{cor1}}
  \ead{huashengxie@gmail.com}
  \author[dlut,veloalpha]{Ruohan Zhang}
  \author[dlut,veloalpha]{Xingyu Li}
  \author[veloalpha]{Feng Zhang}
  \author[dlut]{Zhengxiong Wang}
  \cortext[cor1]{Corresponding author.}
  \address[veloalpha]{Beijing VeloAlpha Technology Co., Ltd., Beijing, 100080, China}
  \address[dlut]{Key Laboratory of Materials Modification by Beams of the Ministry of Education, School of Physics, Dalian University of Technology, Dalian, 116024, China}
  \begin{abstract}
    Tokamak equilibria are commonly exchanged as gridded G-EQDSK files whose conventions, resolutions, and machine-specific formats impede cross-device comparisons and data-driven modeling. Here, we present VEQDB, an open, compact, and reconstructible fixed-boundary equilibrium database built on continuous MXH--Chebyshev geometry and independent physical-profile roots. By decoupling authoritative equilibrium physics from rectangular meshes, VEQDB enables continuous evaluation and metric differentiation at arbitrary application-demanded resolutions. Backed by an automated numerical validation pipeline, VEQDB is structured as an extensible repository for ongoing community expansion. Its inaugural release provides 13,291 accepted equilibria across 267 conventional and spherical tokamaks, encompassing parameter-sampled Grad--Shafranov solutions, G-EQDSK projections spanning EAST, MAST-U, and ITER scales, and controlled variation families with explicit provenance. Benchmark projections reproduce normalized flux maps with RMS errors between $1.09 \times 10^{-3}$ and $1.45 \times 10^{-3}$, while compact JSON representations achieve an 89--96-fold size reduction relative to standard $129 \times 129$ G-EQDSK files. The complete initial release occupies 41~MB in raw JSON and 18~MB in compressed archives, and all records successfully passed independent reload and evaluation tests. VEQDB establishes an extensible, provenance-preserving foundation for equilibrium studies, reduced-order surrogate modeling, and cross-machine workflows.
  \end{abstract}
\end{frontmatter}
\section{Introduction}

Axisymmetric equilibrium reconstruction supplies the magnetic geometry and current profiles used in transport, stability and control studies. EFIT and related methods determine these quantities for individual time slices from device-specific diagnostics~\cite{lao1985efit}. The resulting files retain the conventions of the reconstruction and the source machine. This complicates comparisons across devices and the construction of training sets for reduced-order or data-driven models~\cite{Zheng2025,zhu2020multitokamak}.

Existing infrastructure addresses data access and workflow interoperability, including MDSplus, OMFIT and IMAS~\cite{stillerman1999mdsplus,meneghini2013omfit,imbeaux2015imas}. A complementary requirement is a compact equilibrium representation that separates the physical state from one particular rectangular grid, records how the state was constructed and rejects numerically inadmissible candidates. Such a representation can reduce redundant storage and allow the same equilibrium to be evaluated on a grid chosen by the downstream calculation.

VEQDB addresses this requirement with the MXH--Chebyshev representation~\cite{arbon2021mxh,xie2026mxhchebyshev}. Available experimental or model G-EQDSK references are projected into this state space; direct VEQ solutions provide parameter-sampled equilibria and local variations~\cite{amorisco2024freegsnke,zhang2026veq}. This paper defines the reconstruction and serialization contract, documents a 13,291-record release across 267 devices, and assesses its source fidelity, numerical acceptance, reload reproducibility and storage cost. The emphasis is on a usable fixed-boundary equilibrium resource, not on replacing experimental reconstruction or modelling the exterior vacuum field.

\section{Methods}\label{sec:method}

\subsection{Continuous representation and independent data}

The compact representation separates the continuous equilibrium from any grid on which it is evaluated. Let $r\in[0,1]$ denote a geometric surface label and $\theta$ a periodic poloidal parameter. The magnetic axis and LCFS correspond to $r=0$ and $r=1$, respectively. The MXH map is
\begin{align}
  R(r,\theta)    & = R_0+a\left[h(r)+r\cos\eta(r,\theta)\right],  \\
  Z(r,\theta)    & = Z_0+a\left[v(r)-r\kappa(r)\sin\theta\right], \\
  \eta(r,\theta) & = \theta+c_0(r)+\sum_{m\geq1}
  \left[c_m(r)\cos(m\theta)+s_m(r)\sin(m\theta)\right].
  \label{eq:mxh}
\end{align}
The constants $(R_0,Z_0)$ and $a$ parameterize the boundary reference frame. The functions $h$ and $v$ locate the centers of interior surfaces, $\kappa$ controls elongation, and $c_m$ and $s_m$ distort the poloidal angle. Angular distortion is more economical than an additive Fourier expansion of $R$ and $Z$ because low-order elongation and triangularity are built into the coordinate map~\cite{arbon2021mxh,xie2026mxhchebyshev}.

Each radial shape function has the form
\begin{equation}
  f_m(r)=r^{K_m}\left[f_m^{\mathrm{e}}+(1-r^2)
    \sum_{\ell=0}^{L_m-1}\alpha_{m\ell}T_\ell(2r^2-1)\right],
  \label{eq:radial-basis}
\end{equation}
where $f_m^{\mathrm{e}}$ is its LCFS value and $T_\ell$ is a Chebyshev polynomial. The shift profiles have zero edge values, while $\kappa$, $c_m$, and $s_m$ inherit the independently stored boundary coefficients. For the angular harmonics, $K_m=m$ by default; an optional $K_{\max}$ replaces this by $K_m=\min(m,K_{\max})$. The factor $r^{K_m}$ enforces regularity at the magnetic axis, and $1-r^2$ makes the interior coefficients vanish at the LCFS. Consequently, changing an interior coefficient cannot silently redefine the stored boundary. The native discretization uses $N_r$ Gauss--Lobatto nodes and $N_\theta$ uniformly spaced angles, with spectral radial differentiation and integration. The conversion benchmarks below use $N_r=N_\theta=32$.

Geometry alone is insufficient to reconstruct a physical equilibrium. A record therefore stores four radial roots in addition to the boundary and shape coefficients:
\begin{equation}
  P_\psi=\frac{\dd P}{\dd\psi},\qquad
  FF_\psi=\frac{\dd(F^2/2)}{\dd\psi},\qquad
  \psi_r=\frac{\dd\psi}{\dd r},\qquad
  \psi_{rr}=\frac{\dd^2\psi}{\dd r^2}.
  \label{eq:physical-roots}
\end{equation}
With the native gauge $\psi(0)=0$ and the edge values $P_{\mathrm{e}}$ and $F_{\mathrm{e}}$, the principal profiles follow from
\begin{align}
  \psi(r) & = \int_0^r \psi_r(t)\,\dd t,                             \\
  P(r)    & = P_{\mathrm{e}}-\int_r^1 P_\psi(t)\psi_r(t)\,\dd t,     \\
  F^2(r)  & = F_{\mathrm{e}}^2-2\int_r^1 FF_\psi(t)\psi_r(t)\,\dd t.
  \label{eq:profile-reconstruction}
\end{align}
The sign of $F$ is fixed by the stored toroidal-field reference. The safety factor, toroidal-flux coordinate, current densities, plasma current, surface areas, volumes, and metric coefficients are then derived from Eqs.~\eqref{eq:mxh}--\eqref{eq:profile-reconstruction}. They are not independent fitting inputs and are not stored as second authoritative copies.

The compact JSON schema mirrors this ownership boundary. It contains the reconstruction roots, their discrete configuration, the LCFS data, and the integration constants. Human-facing quantities such as $(R_{\mathrm{axis}},Z_{\mathrm{axis}})$, $q_0$, $q_{\min}$, $q_{95}$, $I_p$, $\beta_t$, and stored thermal energy are included as diagnostics for inspection. A reader ignores these diagnostic values and reconstructs them from the roots. This rule allows future diagnostic definitions to change without changing the meaning of an archived equilibrium. The compact JSON field reference in \ref{app:json-fields} gives the exact output order, field roles, units, and omission rules of the current schema.

\subsection{Two numerical construction routes}

VEQDB uses two numerical routes. VEQ generates an equilibrium from prescribed boundary and physical constraints; this route covers both the device-parameter survey and controlled variations around a reference. Joint fitting converts an existing G-EQDSK equilibrium into the same compact representation. Both routes supply the geometry and physical profiles required for reconstruction, but their provenance and evidential roles remain distinct.

\subsubsection{Direct fixed-boundary solution}

VEQ solves the axisymmetric Grad--Shafranov equation
\begin{equation}
  \Delta^*\psi+FF_\psi+\mu_0R^2P_\psi=0,
  \qquad
  \Delta^*\psi=R\frac{\partial}{\partial R}
  \left(\frac{1}{R}\frac{\partial\psi}{\partial R}\right)
  +\frac{\partial^2\psi}{\partial Z^2},
  \label{eq:gs}
\end{equation}
directly on the finite MXH--Chebyshev manifold~\cite{zhang2026veq}. The LCFS is prescribed, whereas the interior geometry and the route-dependent profile coefficients are nonlinear unknowns. After transformation to $(r,\theta)$, VEQ projects the signed force-balance defect onto test functions induced by admissible variations of the representation. The resulting algebraic residual is a variationally structured projection of Eq.~\eqref{eq:gs}; it is not a distance fit to a reference flux map.

The accepted solution directly supplies the geometry and profiles needed by the compact format, so serialization requires no intermediate grid fit. The formulation and solver are described in Ref.~\cite{zhang2026veq}. Convergence of the projected residual establishes a solution within the selected coefficient space; assessing representation error additionally requires a refined or independent reference.

\subsubsection{Joint projection of a G-EQDSK file}

The second route treats a G-EQDSK file as reconstructed or model-generated input. It consumes the rectangular flux map $\psi_g(R,Z)$, the reported magnetic axis and LCFS, the signed axis-to-edge flux interval, and the tabulated $P_\psi$ and $FF_\psi$ profiles. The conversion first verifies that the axis lies inside the LCFS and that a nested sequence of closed contours can be extracted from the grid. Independent contour fits provide phase-aligned estimates of the surface shapes. These estimates initialize the nonlinear solve; they do not define the final boundary or enter the accepted record as a separate source of truth.

To preserve the file's flux orientation, the interpolated target is normalized as
\begin{equation}
  s_g(R,Z)=\frac{\psi_g(R,Z)-\psi_{\mathrm{axis}}}
  {\psi_{\mathrm{LCFS}}-\psi_{\mathrm{axis}}}.
  \label{eq:file-flux}
\end{equation}
The fitted map uses
\begin{equation}
  s_\beta(r)=r^2\left[1+(1-r^2)\sum_{\ell=0}^{L_\psi-1}
    \beta_\ell T_\ell(2r^2-1)\right],
  \label{eq:flux-map}
\end{equation}
which satisfies $s_\beta(0)=0$ and $s_\beta(1)=1$ identically and has the regular leading behavior $s_\beta\sim r^2$ at the axis. The unknown vector contains the active LCFS parameters $b$, the interior shape coefficients $u$, and the flux coefficients $\beta$. The magnetic-axis coordinates are held exactly by eliminating the constant terms of $h$ and $v$ after every boundary update.

The final objective combines the interior flux map with a direct LCFS constraint:
\begin{equation}
  \mathcal{L}(b,u,\beta)=
  \frac{1}{N_f}\sum_{i,j}e_{ij}^2
  +\frac{w}{N_b}\sum_{k}
  \left(\frac{2d_k(b)}{a_{\mathrm{init}}}\right)^2,
  \quad
  e_{ij}=s_g\!\left(X_{ij}(b,u)\right)-s_\beta(r_i),
  \label{eq:joint-loss}
\end{equation}
where $X_{ij}=(R(r_i,\theta_j),Z(r_i,\theta_j))$, $d_k$ is the distance obtained by local closest-point projection from a file-boundary point to the candidate LCFS, and $w=1$ in the maintained implementation. Each block is normalized by its own sample count. The scale $2/a_{\mathrm{init}}$ converts boundary displacement to the order of the flux change expected from the circular reference $s\simeq r^2$. It is fixed at initialization, so the two terms retain the same meaning while $a$ changes. Neither this scale nor $w$ represents a measured noise variance.

The LCFS term resolves a specific degeneracy of a flux-only objective. For a small normal displacement $\delta n$ away from a critical point,
\begin{equation}
  \delta s_g\simeq |\nabla s_g|\,\delta n.
  \label{eq:flux-distance}
\end{equation}
Thus, the interior term in Eq.~\eqref{eq:joint-loss} is locally a normal-distance fit weighted by $|\nabla s_g|^2$. Near an X point, $|\nabla s_g|$ becomes small and a visibly displaced boundary can contribute little flux error. The continuous-distance block restores a direct positional constraint there without replacing the flux information in the plasma interior.

The implementation evaluates $s_g$ and its $R$ and $Z$ derivatives by bilinear interpolation. Its analytic Jacobian contains the two essential chains
\begin{equation}
  \frac{\partial e_{ij}}{\partial u_p}
  =\nabla s_g(X_{ij})\mathbin{\cdot}
  \frac{\partial X_{ij}}{\partial u_p},
  \qquad
  \frac{\partial e_{ij}}{\partial\beta_\ell}
  =-r_i^2(1-r_i^2)T_\ell(2r_i^2-1).
  \label{eq:fit-jacobian}
\end{equation}
At every trial boundary, the LCFS samples are reprojected onto the continuous curve. The normal derivative of the projection residual removes the arbitrary tangential parameterization from the boundary update. Interior residuals use the $N_r$ Lobatto radii and $\max(64,2N_\theta)$ poloidal angles; the boundary block uses at least the same angular resolution and explicitly includes geometric extrema.

Columns of the full Jacobian are scaled before the Levenberg--Marquardt step,
\begin{equation}
  \left(\widehat J^{\mathsf T}\widehat J+\lambda I\right)\delta
  =-\widehat J^{\mathsf T}e.
  \label{eq:lm}
\end{equation}
The damping parameter is updated between accepted and rejected steps. Iteration
stops when the relative loss decrease is at most $10^{-6}$ or when the infinity
norm of the scaled gradient is at most
$10^{-8}\max(\sqrt{\mathcal L},\epsilon_{\mathrm{mach}})$.

Numerical termination produces a candidate rather than an equilibrium. Acceptance separately requires stationary and locally minimizing LCFS projections, a full-column-rank scaled Jacobian, and positive finite $R$, $\kappa$, geometric Jacobian, and $\dd s_\beta/\dd r$ on a denser grid excluding the magnetic axis. The physical reconstruction must also produce finite magnetic fields, pressures, currents, safety factors, surface coefficients, areas, and volumes. Only then is an immutable equilibrium returned. Initialization, nonlinear solution, projection, rank, geometry, flux, and physical-reconstruction failures remain distinguishable in the reported error. No equilibrium instance is returned if any acceptance check fails.

The accepted conversion preserves the signed flux interval, $P_{\mathrm e}$, $F_{\mathrm e}$, and the file information in $P_\psi$ and $FF_\psi$. The native flux gauge is reset to $\psi(0)=0$ without changing the axis-to-edge interval. The pressure and $F$ profiles are integrated on the native Lobatto grid. The safety factor and total plasma current are reconstructed afterward and are not fitted targets. G-EQDSK conversion is consequently a projection of file data onto the compact representation, whereas VEQ solves a projected physical equation.

\section{Representation accuracy and software performance}\label{sec:performance}

\subsection{G-EQDSK conversion accuracy and cost}

Conversion accuracy is assessed separately for the sampled flux, boundary geometry, and reconstructed plasma current. These measures quantify agreement with the input file. A force-balance defect already present in that file may remain after fitting, because Eq.~\eqref{eq:joint-loss} imposes data agreement rather than the Grad--Shafranov equation.

Table~\ref{tab:geqdsk} reports conversion of a CHEASE H-mode equilibrium and an EFIT X-point equilibrium at four coefficient budgets. Following the VEQ convention~\cite{zhang2026veq}, the number in parentheses is the active internal MXH--Chebyshev parameter count. The flux-map RMS is evaluated by interpolating the file onto the accepted $(R,Z)$ nodes. The LCFS column reports the one-sided RMS distance from each file boundary point to the fitted boundary. Plasma current is reconstructed from the accepted equilibrium and compared with the value reported in the file.

\begin{table}[H]
  \centering
  \caption{G-EQDSK conversion at $N_r=N_\theta=32$. Coefficient time is a warm five-run median on an Apple M4 Pro and excludes first compilation, contour extraction, acceptance, and object construction.}
  \label{tab:geqdsk}
  \begin{tabular}{lrrrr}
    \toprule
    Case (parameters) & Flux-map RMS (--)    & LCFS RMS (mm) & $\Delta I_p/I_{p,g}$ (\%) & Coefficient time (ms) \\
    \midrule
    H-mode (29)       & $3.470\times10^{-3}$ & 0.726         & $-0.852$ & 3.23 \\
    H-mode (41)       & $2.354\times10^{-3}$ & 0.729         & $-0.217$ & 4.08 \\
    H-mode (61)       & $7.978\times10^{-4}$ & 0.714         & $+0.134$ & 7.13 \\
    H-mode (130)      & $4.953\times10^{-4}$ & 0.714         & $+0.086$ & 12.54 \\
    X-point (21)      & $8.994\times10^{-3}$ & 0.811         & $-0.127$ & 8.06 \\
    X-point (31)      & $4.378\times10^{-3}$ & 0.610         & $-0.017$ & 5.25 \\
    X-point (96)      & $1.143\times10^{-3}$ & 0.645         & $+0.002$ & 7.49 \\
    X-point (130)     & $8.433\times10^{-4}$ & 0.653         & $+0.006$ & 10.30 \\
    \bottomrule
  \end{tabular}
\end{table}

All eight joint fits were accepted. From the smallest to the 130-parameter configuration, the normalized flux-map RMS decreased by factors of 7.0 for H-mode and 10.7 for X-point. The derived current difference fell below $0.1\%$ in both 130-parameter cases. The LCFS distance changed much less because its explicit residual block was already well resolved at the smaller budgets. Neither the LCFS error nor every derived quantity is required to decrease monotonically with parameter count: the two residual blocks compete under a fixed weight, and the derived quantities are outside the fitting objective. The reported accuracy must therefore retain separate geometry, flux, force-balance, and reconstructed-physics measures\cite{lee2015ecom}.

\subsection{Encoding, size, and processing cost}

G-EQDSK tabulates a rectangular $\psi(R,Z)$ map, profiles and boundary samples,
whereas VEQDB stores the coefficients needed to reconstruct the fixed-boundary
plasma interior. Its authoritative size is therefore independent of an evaluation
grid. For the three cross-device examples, ten-significant-digit compact JSON is
89--96 times smaller than the $129\times129$ source files (Table~\ref{tab:geqdsk-storage}).
The same representation becomes more favourable as the source grid is refined,
because it does not store the rectangular flux map.

\begin{table}[H]
  \centering
  \caption{File size after projecting representative G-EQDSK inputs onto the compact fixed-boundary representation. The $257\times257$ MAST-U and ITER entries use the corresponding high-resolution FreeGSNKE files; the compact representation remains independent of the source-grid resolution.}
  \label{tab:geqdsk-storage}
  \begin{tabular}{lrrr}
    \toprule
    Case   & G-EQDSK (kB) & Compact JSON (kB) & Reduction \\
    \midrule
    EAST ($129\times129$)   & 285.49 & 3.16              & $90.4\times$ \\
    MAST-U ($129\times129$) & 290.23 & 3.24              & $89.5\times$ \\
    ITER ($129\times129$)   & 298.00 & 3.11              & $95.9\times$ \\
    MAST-U ($257\times257$) & 1110.99 & 3.24              & $342.9\times$ \\
    ITER ($257\times257$)   & 1133.55 & 3.11              & $364.5\times$ \\
    \bottomrule
  \end{tabular}
\end{table}

This is not lossless G-EQDSK compression: the compact record contains the LCFS
and plasma interior but omits the exterior rectangular flux map. The coefficients
can nevertheless be evaluated and differentiated on any selected grid.

For a representative $32\times32$ equilibrium with 100 active geometry
coefficients, 254 independent float64 values occupy 2032~B; configuration and
format metadata raise the uncompressed reconstruction reference to 2078~B.
Table~\ref{tab:encoding} compares encodings of the same state.

\begin{table}[t]
  \centering
  \caption{Storage for a representative $32\times32$ equilibrium with 100 active geometry coefficients.}
  \label{tab:encoding}
  \begin{tabular}{lrr}
    \toprule
    Encoding                  & Size (B) & Reference (\%) \\
    \midrule
    Compact JSON, float64     & 5555     & 267.3          \\
    Compact JSON, 10 digits   & 3928     & 189.0          \\
    Compact Pickle            & 2881     & 138.6          \\
    Binary, Zstandard level 9 & 1836     & 88.4           \\
    \bottomrule
  \end{tabular}
\end{table}

The reference size is an accounting value, not an information-theoretic bound or
the memory use of a live object. Compact JSON retains readable names and decimal
values; binary uses integer differencing, byte reordering and Zstandard level-9
compression, reaching 1836~B because the coefficients are compressible. Compact
JSON remains the exchange format because its 2--6~kB records are directly
inspectable and language independent; binary and Pickle serve bulk archival and
trusted Python workflows, respectively.

After warm-up, compact-JSON export, import and file-cache reads took 129--188,
58--81 and 77--99~$\mu$s, respectively. These timings exclude equilibrium
generation and first compilation. Coefficient fitting required 3.2--12.5~ms
(Table~\ref{tab:geqdsk}); complete conversion, including contour extraction,
acceptance and physical reconstruction, required approximately 17--32~ms.

\section{Database content and quality assessment}\label{sec:database}

VEQDB contains 13,291 accepted conventional- and spherical-tokamak equilibria
from 267 devices in a common compact representation. It packages the
representation, released records and fixed-boundary VEQ workflow.
Parameter-sampled and local-variation records use direct VEQ solutions;
reference-derived records are projections of retained G-EQDSK files. Provenance
separates experimental reconstructions, public machine models and synthetic
samples, which have different scientific uses.

\begin{figure*}[!htbp]
\centering
\includegraphics[width=\doublefigurewidth]{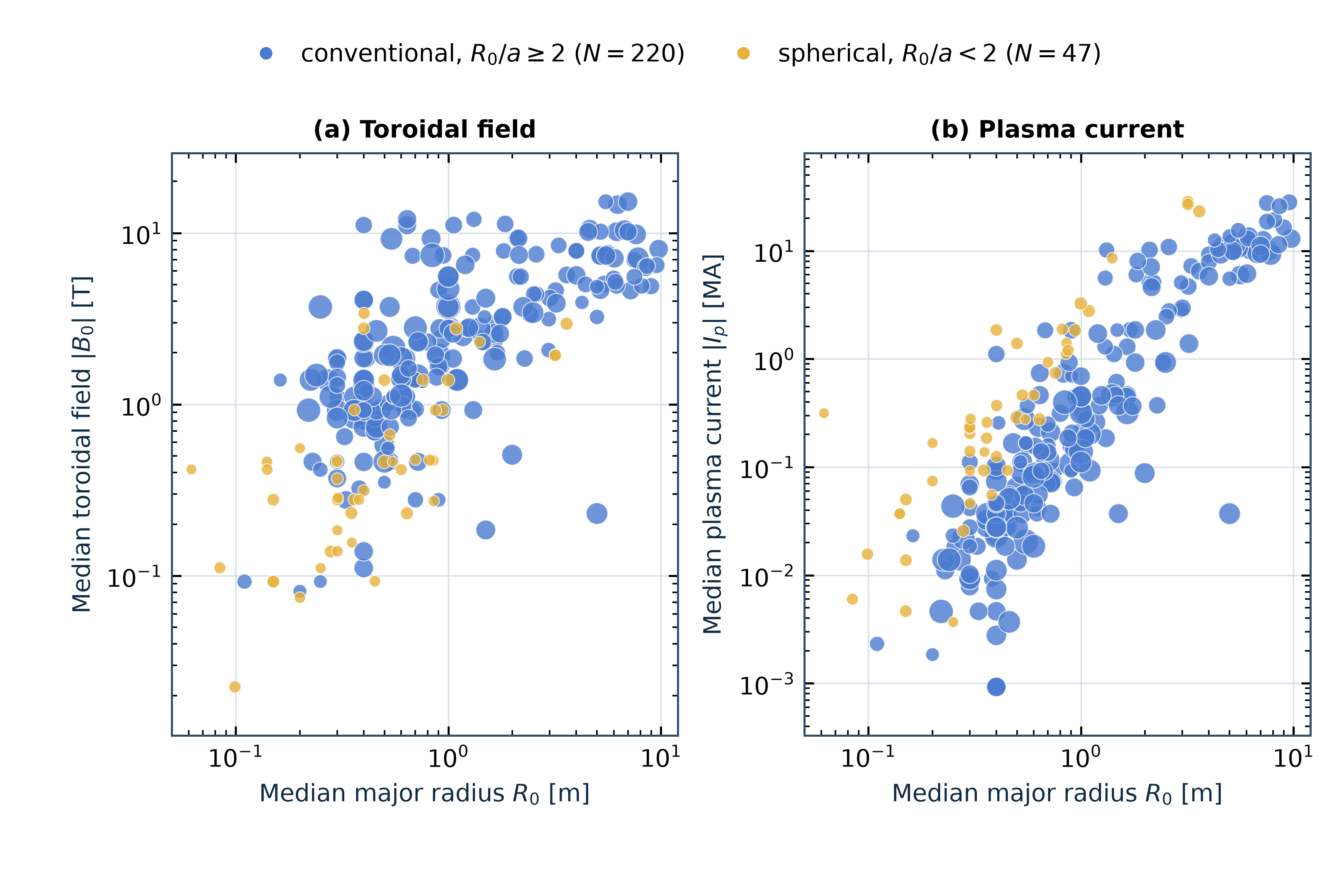}
\caption{Device-scale coverage of the final release. Each marker represents one
of the 267 devices at the median values of its accepted records. Colour separates
conventional and spherical tokamaks at $R_0/a=2$, and marker area is directly
proportional to the median $R_0/a$. The panels show (a) toroidal field and
(b) plasma current against major radius on logarithmic axes. The point cloud
describes the coverage of the released records rather than an experimental
operating distribution.}
\label{fig:device-coverage}
\end{figure*}

Figure~\ref{fig:device-coverage} summarises the device-scale coverage of the
released equilibria. Each point uses the median $R_0$, $a$, $|B_0|$, and $|I_p|$
of the accepted records for one device. The release spans more than two orders of
magnitude in major radius and nearly five orders of magnitude in plasma current.
Marker area is proportional to the median aspect ratio $R_0/a$. These medians
describe the generated equilibrium set rather than experimental operating
distributions.

Table~\ref{tab:representative-devices} gives a compact view of the release. It
lists familiar devices rather than attempting an encyclopaedic inventory; the
reported values are medians of the accepted records in this release and do not
represent experimental operating points.

\begin{table}[!htbp]
  \centering
  \caption{Representative devices in VEQDB. The release contains 267 devices and 13,291 accepted equilibria in total.}
  \label{tab:representative-devices}
  \begin{tabular}{lrrrr}
    \toprule
    Device & $R_0$ (m) & $a$ (m) & $|B_0|$ (T) & $|I_p|$ (MA) \\
    \midrule
    EAST & 1.80 & 0.450 & 3.24 & 0.924 \\
    DIII-D & 1.70 & 0.600 & 2.00 & 1.86 \\
    ASDEX Upgrade & 1.65 & 0.502 & 2.97 & 1.29 \\
    TCV & 0.880 & 0.253 & 1.44 & 0.927 \\
    NSTX-U & 0.941 & 0.546 & 0.927 & 1.84 \\
    KSTAR & 1.80 & 0.499 & 3.24 & 1.85 \\
    JT-60SA & 2.96 & 1.18 & 2.07 & 5.09 \\
    SPARC & 1.85 & 0.569 & 11.3 & 8.04 \\
    ITER & 6.20 & 2.00 & 4.93 & 13.8 \\
    $\cdots$ & $\cdots$ & $\cdots$ & $\cdots$ & $\cdots$ \\
    \bottomrule
  \end{tabular}
\end{table}

\subsection{Parameter-sampled equilibria}\label{sec:ch3-parameterised}

The largest part of the release is constructed from device parameters rather than
from individual experimental reconstructions. Candidate equilibria are drawn by
scrambled low-discrepancy sampling over the minor radius, edge elongation and
triangularity, radial boundary displacement, field and current magnitudes,
toroidal beta, and four GAQ profile exponents. The source metadata retain the
origin of each device parameter and flag estimated shape parameters. The two
representative batches considered below use $N_r=N_\theta=32$ and vary the field
and current magnitudes by factors of $0.85$--$1.00$ relative to their device
references.

\begin{figure}[!htbp]
\centering
\includegraphics[width=0.62208\textwidth]{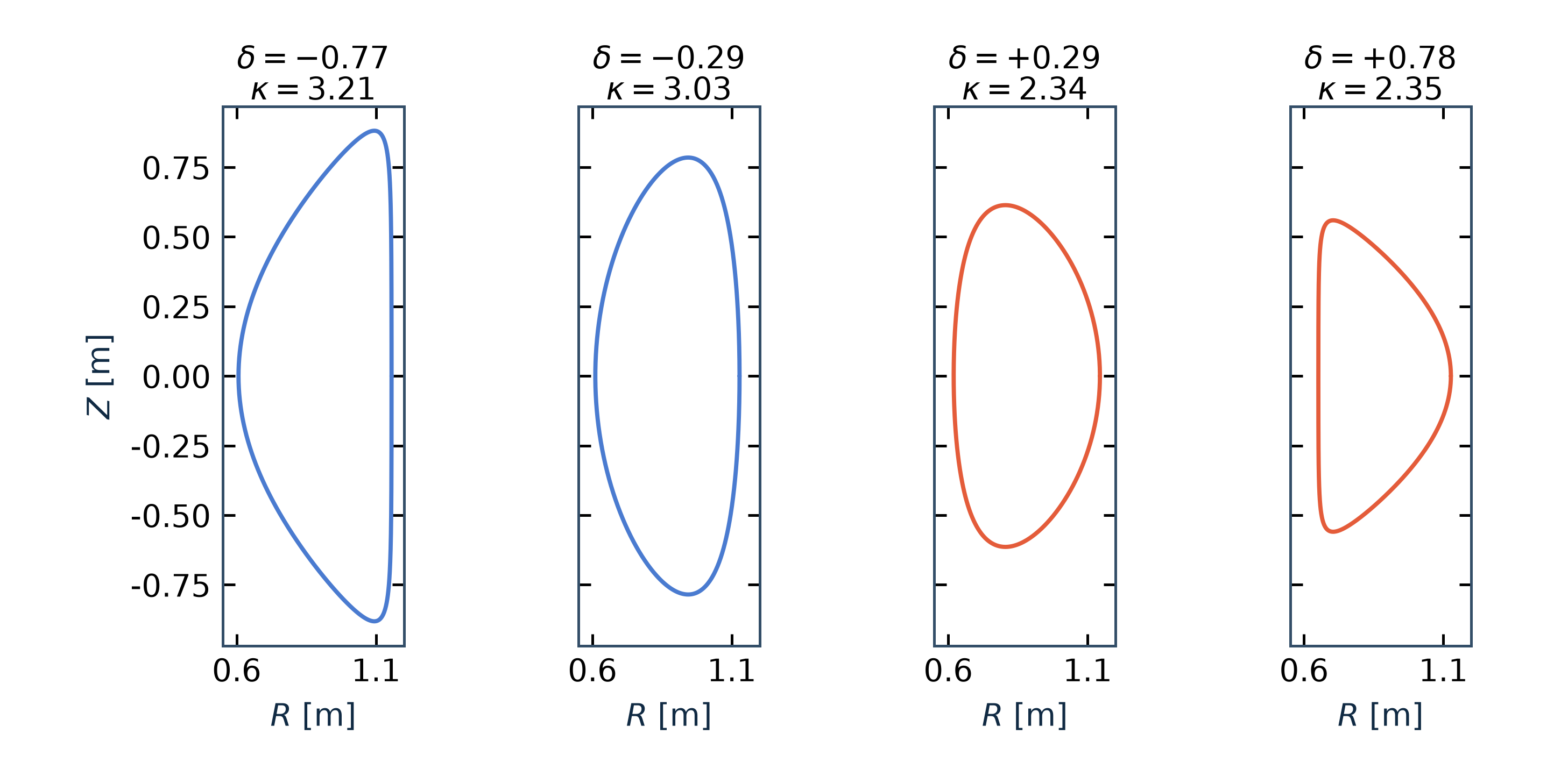}
\caption{Four last-closed-flux surfaces from the 50-case TCV batch, selected at
the endpoints and approximately one-third and two-thirds positions of the sorted
sampled triangularity. Each case occupies one column; the two-line title gives
the sampled triangularity and elongation. The shared vertical axis permits direct
shape comparison. Size and radial displacement vary as well, so the cases are
not a single-factor scan.}
\label{fig:ch3-geometry}
\end{figure}

Table~\ref{tab:ch3-batches} summarises the sampled ranges and convergence
statistics of the two 50-case batches.
Figure~\ref{fig:ch3-geometry} makes the geometric coverage explicit by placing
four TCV boundaries side by side. The cases are selected at the endpoints and the
approximate thirds of the sorted sampled triangularity, giving two negative- and
two positive-triangularity examples. Elongation, size, and radial position vary
with triangularity. These are synthetic fixed-boundary configurations, so the figure
demonstrates coverage of the sampled range rather than an experimental operating
envelope, and because several parameters move together it is not a
single-parameter causality test. The imposed boundary centre,
$R_{\mathrm c}=R_{0,\mathrm{ref}}+\Delta R_{\mathrm b}$, is distinguished from the
solved magnetic-axis position, so a change in $R_{\mathrm{axis}}-R_{\mathrm c}$
can be separated from the prescribed translation of the boundary.

\begin{table}[!htbp]
\centering
\caption{Measured coverage and convergence of the two parameter-sampled batches.
Geometric, field and current ranges are those of the accepted records; the two
residual rows are the projected residuals of the $N_r=N_\theta=32$ solve, which
measure convergence inside the selected coefficient space rather than agreement
with an independent reference.}
\label{tab:ch3-batches}
\begin{tabular}{lrr}
\toprule
Metric & START & TCV\\
\midrule
Accepted cases & 50 & 50\\
$a$ (m) & 0.207--0.253 & 0.226--0.275\\
$\kappa_{\mathrm e}$ & 1.441--2.145 & 2.303--3.344\\
$\delta_{\mathrm e}$ & 0.305--0.697 & $-0.775$--$0.777$\\
$B_t$ (T) & 0.264--0.309 & 1.310--1.538\\
$|I_p|$ (MA) & 0.257--0.299 & 0.852--0.998\\
$\beta_t$ (\%) & 5.18--19.66 & 0.573--2.994\\
$P_{\mathrm{axis}}$ (kPa) & 6.14--27.80 & 15.71--142.01\\
Residual median & $7.16\times10^{-11}$ & $1.41\times10^{-10}$\\
Residual maximum & $7.36\times10^{-10}$ & $1.19\times10^{-9}$\\
\bottomrule
\end{tabular}
\end{table}

The source profiles of these equilibria are not tabulated per case. They are
generated from the GAQ shape functions
\begin{equation}
 g_p(x)=(1-x^{n_p})^{m_p},\qquad
 g_f(x)=(1-x^{n_f})^{m_f},\qquad
 x=\hat{\psi}=\frac{\psi-\psi_{\mathrm{axis}}}{\Delta\psi},
 \label{eq:ch3-gaq-shapes}
\end{equation}
with $\Delta\psi=\psi_{\mathrm e}-\psi_{\mathrm{axis}}$ retaining its sign. The
two batches sample $n_p\in[1,2.5]$, $m_p\in[1,2]$,
$n_f\in[1.5,3]$, and $m_f\in[1,1.8]$. At fixed $n$, raising $m$ concentrates the
normalized source toward the core; at fixed $m$, raising $n$ keeps the core
broader before the edge decay sets in. The values are evaluated analytically on
normalized-flux nodes and interpolated by the library during the solve. Their
common normalization is fixed by the total current and their relative weighting is
adjusted until the target beta is reached, so the two source amplitudes are not
independent free parameters. Written with the amplitudes of the accepted solution,
\begin{equation}
 \frac{\mathrm dP}{\mathrm dx}=-C_p g_p(x),\qquad
 \frac{\mathrm d(F^2/2)}{\mathrm dx}=-C_f g_f(x),
 \qquad
 P(x)=C_p\int_x^1g_p(u)\,\mathrm du,
 \label{eq:ch3-gaq-sources}
\end{equation}
where $C_p,C_f>0$ carry units of Pa and $\mathrm{T^2\,m^2}$, respectively, and
$P(1)=0$ holds by construction. The derivatives used in Sec.~\ref{sec:method} obey
$P_\psi=(\mathrm dP/\mathrm dx)/\Delta\psi$ and
$FF_\psi=[\mathrm d(F^2/2)/\mathrm dx]/\Delta\psi$. Quoting $C_p$ and $C_f$
rather than implementation-specific mixture coefficients keeps the physical
amplitude separate from the normalization conventions of a particular generator
version; the full parameter set and its normalization are retained in each input
record. The naming follows the GAQ free-boundary code family, which introduced a
library of standard current-density profiles written as $J^*\propto R\,G(\psi)$ with
$G$ taking a linear, polynomial or power form selected by the user
\cite{mcclain1977gaq}. The parameterization used here is that power family carried
over to the normalized flux of a fixed-boundary equilibrium, with the two exponents
per source and the amplitudes $C_p$ and $C_f$ fixed as described above; the exact
coefficients and normalization of every case are stored with it rather than
inferred from this text.

\begin{figure*}[!t]
\centering
\includegraphics[width=\doublefigurewidth]{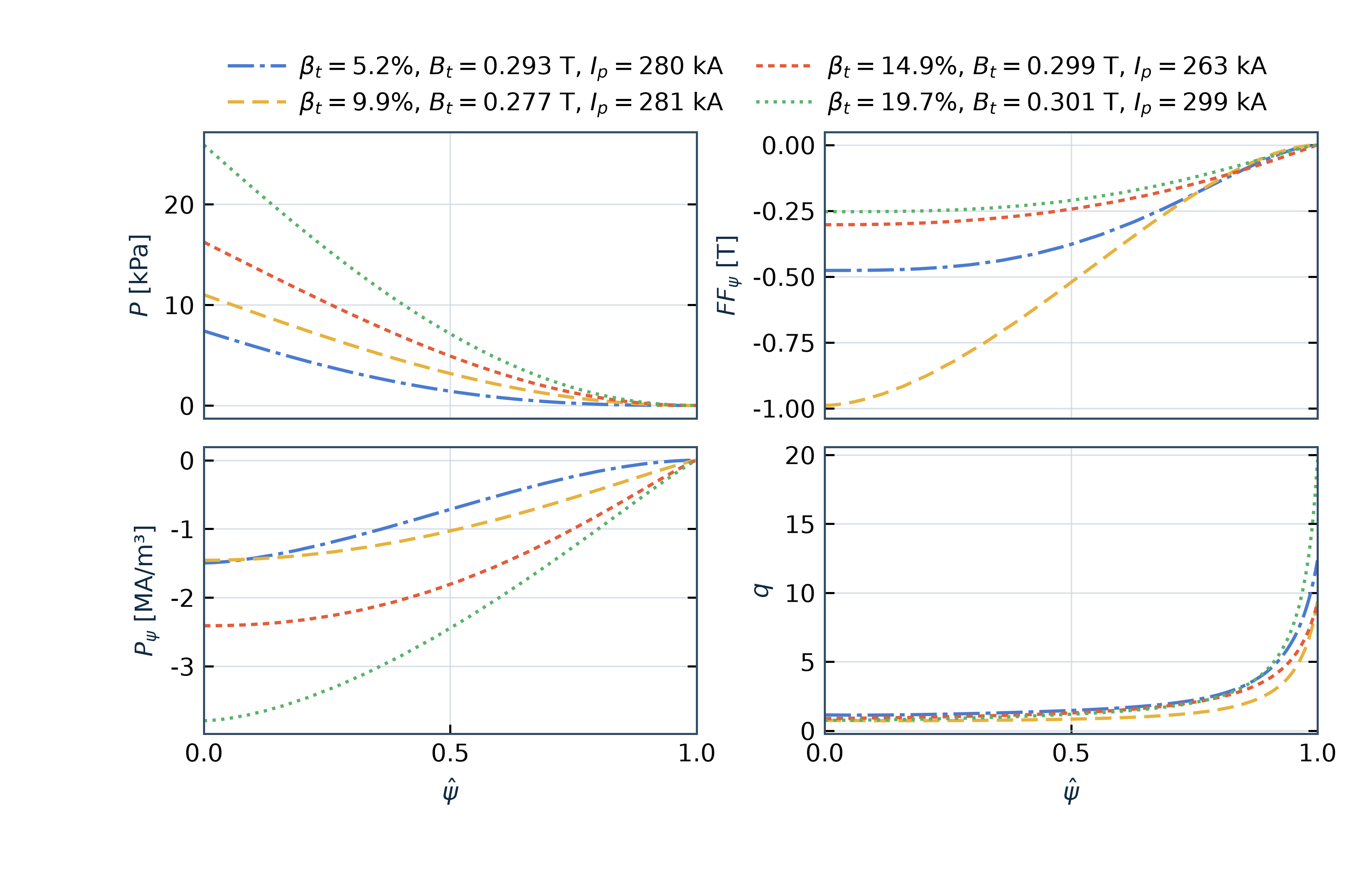}
\caption{Four START cases selected nearest to $\beta_t=5\%$, $10\%$, $15\%$, and
$20\%$; the realized values are $5.2\%$, $9.9\%$, $14.9\%$, and $19.7\%$. From
upper left to lower right, the panels show $P$, $FF_\psi$, $P_\psi$, and $q$.
Within each column, the two panels share one $\hat{\psi}$ axis and one set of
ticks. The legend gives the realized beta, toroidal field, and current of each
sample. Geometry and source parameters vary jointly, and the curves are drawn at a
raised display resolution ($192\times128$); all reported values in this section
remain those of the $32\times32$ stored solution.}
\label{fig:ch3-profiles}
\end{figure*}

Figure~\ref{fig:ch3-profiles} compares four START cases selected nearest to
$\beta_t=5\%$, $10\%$, $15\%$, and $20\%$, deliberately retaining the
high-$\beta_t$ end of the batch as an illustrative spherical-tokamak case. Because
the sampled shapes and source exponents move together, $P$, $FF_\psi$, $P_\psi$,
and the reconstructed $q$ profiles differ in both magnitude and interior shape.
Here $P_\psi=\dd P/\dd\psi$ is evaluated with the native signed poloidal flux.
For these START cases $\Delta\psi>0$ and pressure decreases toward the LCFS, so
$P_\psi$ is negative. The corresponding on-axis pressures span
$6.14$--$27.80$~kPa for START and $15.71$--$142.01$~kPa for TCV, evaluated with
\begin{equation}
 \beta_t=\frac{2\mu_0\langle P\rangle_V}{B_t^2},\qquad
 \langle P\rangle_V=\frac{1}{V}\int_V P\,\mathrm dV .
 \label{eq:ch3-beta}
\end{equation}
The spherical-tokamak batch reaches higher toroidal beta while its on-axis
pressure range is lower than that of the conventional batch. Its toroidal field
is approximately five times smaller, so the observed beta ranges are consistent
with Eq.~\eqref{eq:ch3-beta}; because geometry and source profiles vary jointly,
the two batches do not isolate a field-only causal effect. Reporting $B_t$, $I_p$,
pressure, and beta together makes this distinction explicit. These ranges describe
accepted equilibria of a sampling procedure; they do not define an MHD stability
limit.

\subsection{Reference-derived equilibria}\label{sec:ch3-gfile}

The reference-derived subset comprises an experimental EAST EFIT reconstruction
\cite{gtaw_gfiles,lao1985efit} and model equilibria obtained with FreeGSNKE for
MAST-U and ITER-scale machine configurations~\cite{amorisco2024freegsnke}. The
EAST source is a $129\times129$ lower-single-null reconstruction at
$B_t=1.85$~T and $I_p=0.352$~MA; its flux map contains a separatrix-level saddle
at $(R,Z)=(1.630,-0.765)$~m. The MAST-U and ITER cases test the representation at
smaller and larger machine scales, but are labelled as model equilibria rather
than experimental reconstructions. For the ITER case, the open separatrix reaches
the source-domain boundary; conversion therefore uses the closed inset contour at
$\hat{\psi}=0.999$.

Re-solving the EAST case on the fitted boundary with the fitted source profiles
gives a projected residual of $1.0\times10^{-11}$ and a $0.38$~mm boundary RMS
difference. This test establishes convergence within the VEQ coefficient space;
profile agreement with the G-EQDSK input remains a separate conversion metric.

\begin{figure*}[!htbp]
\centering
\includegraphics[width=\doublefigurewidth]{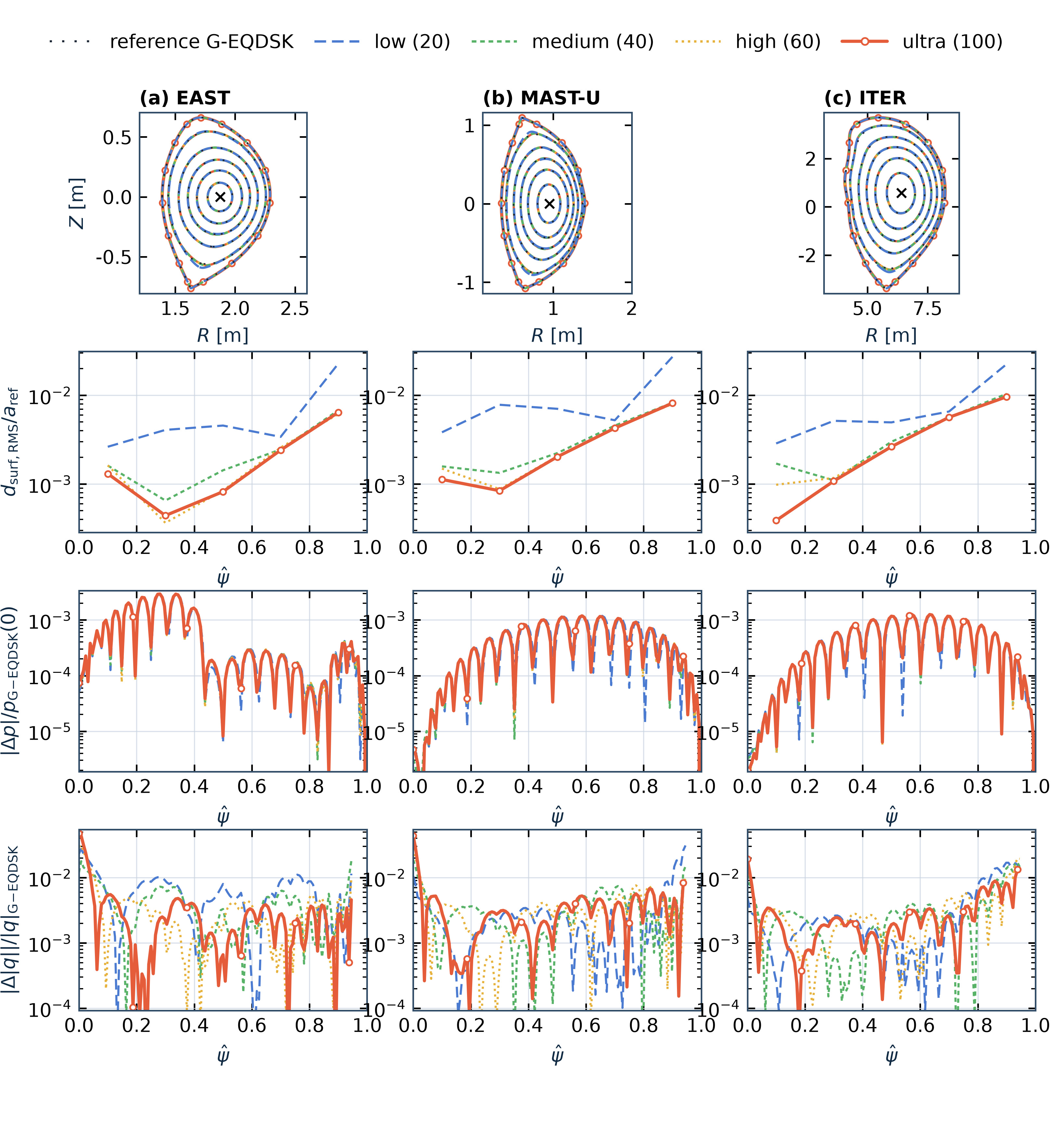}
\caption{Cross-device comparison for EAST discharge 48059 at 3.65~s, a MAST-U
inverse-solve equilibrium and an ITER inverse-solve equilibrium. The top row
shows source G-EQDSK magnetic surfaces and boundary in black, with compact fits
at 20, 40, 60 and 100 active interior parameters in colour. The second row shows
the RMS closest-curve distance from each fitted flux surface to the corresponding
G-EQDSK surface, normalized by the reference minor radius. The third and fourth
rows show $|p_{\mathrm{fit}}-p_{\mathrm{G\! -\! EQDSK}}|/p_{\mathrm{G\! -\! EQDSK}}(0)$
and $\big||q|_{\mathrm{fit}}-|q|_{\mathrm{G\! -\! EQDSK}}\big|/|q|_{\mathrm{G\! -\! EQDSK}}$,
respectively, for the same fits on logarithmic axes. All fits use the same 15th-order cosine
and sine LCFS parameterization. The surfaces are drawn at
$\hat{\psi}=0.1,0.3,0.5,0.7,0.9$, and the $q$ panels stop at
$\hat{\psi}=0.95$ because the separatrix value is ill-conditioned. All inputs
are canonicalised before fitting so that $I_p>0$ and $\psi_{\rm lcfs}>0$.}
\label{fig:ch3-gfile-compare}
\end{figure*}

Figure~\ref{fig:ch3-gfile-compare} tests whether increasing the interior fitting
budget materially improves the reconstructed magnetic surfaces and profiles. Before fitting,
each G-EQDSK input is mapped to the comparison convention with positive $I_p$ and an
outward-increasing poloidal-flux interval. The same transformation is applied to
$\psi$, $P_\psi$ and $FF_\psi$, so the normalized flux map is unchanged. This
produces compact records with positive \code{Ip} and \code{psi_lcfs}. The profile
lower rows compare the fitted $p$ and $|q|$ directly with the source G-EQDSK input, while the
plasma current is compared as a signed scalar diagnostic.

The four budgets in Fig.~\ref{fig:ch3-gfile-compare} contain 20, 40, 60 and 100
active interior fitting parameters, respectively. All fits retain the same
15th-order cosine and sine LCFS parameterization and the same $32\times32$
materialization. Increasing the interior budget does not produce a monotonic change in
the signed current error (Table~\ref{tab:ch3-current}), and the magnetic surfaces
are already visually close at the lowest budget. This plateau is an empirical
property of these source files and fitting settings. It may reflect the
non-orthogonal, nonlinear MXH parameterization and the competing flux and LCFS
terms in the joint objective, but the present comparison does not establish a
general approximation limit.

\clearpage

\begin{table}[H]
\centering
\caption{Signed plasma-current difference, $100(I_{p,\rm fit}-I_{p,\rm ref})/I_{p,\rm ref}$, after canonicalising each G-EQDSK input to positive $I_p$. The columns correspond to 20, 40, 60 and 100 active interior fitting parameters; the LCFS parameterization is fixed across the four fits.}
\label{tab:ch3-current}
\begin{tabular}{lrrrrr}
\toprule
Reference & $I_{p,\rm ref}$ (MA) & 20 (\%) & 40 (\%) & 60 (\%) & 100 (\%)\\
\midrule
EAST & 0.3524 & $-0.1162$ & $-0.0723$ & $-0.0860$ & $-0.0889$\\
MAST-U & 0.6000 & $-0.1154$ & $-0.0570$ & $-0.0579$ & $-0.0594$\\
ITER & 15.0000 & $+2.4375$ & $+2.4356$ & $+2.4321$ & $+2.4294$\\
\bottomrule
\end{tabular}
\end{table}

\subsection{Local variations of a converted reference}\label{sec:ch3-variation}

A converted reference can seed local variations because boundary and source
profiles are independent components of the stored state. Perturbing these
components and re-solving the fixed-boundary problem samples a neighbourhood of
the reference without repeating the G-EQDSK projection.

\begin{figure*}[!htbp]
\centering
\includegraphics[width=\doublefigurewidth]{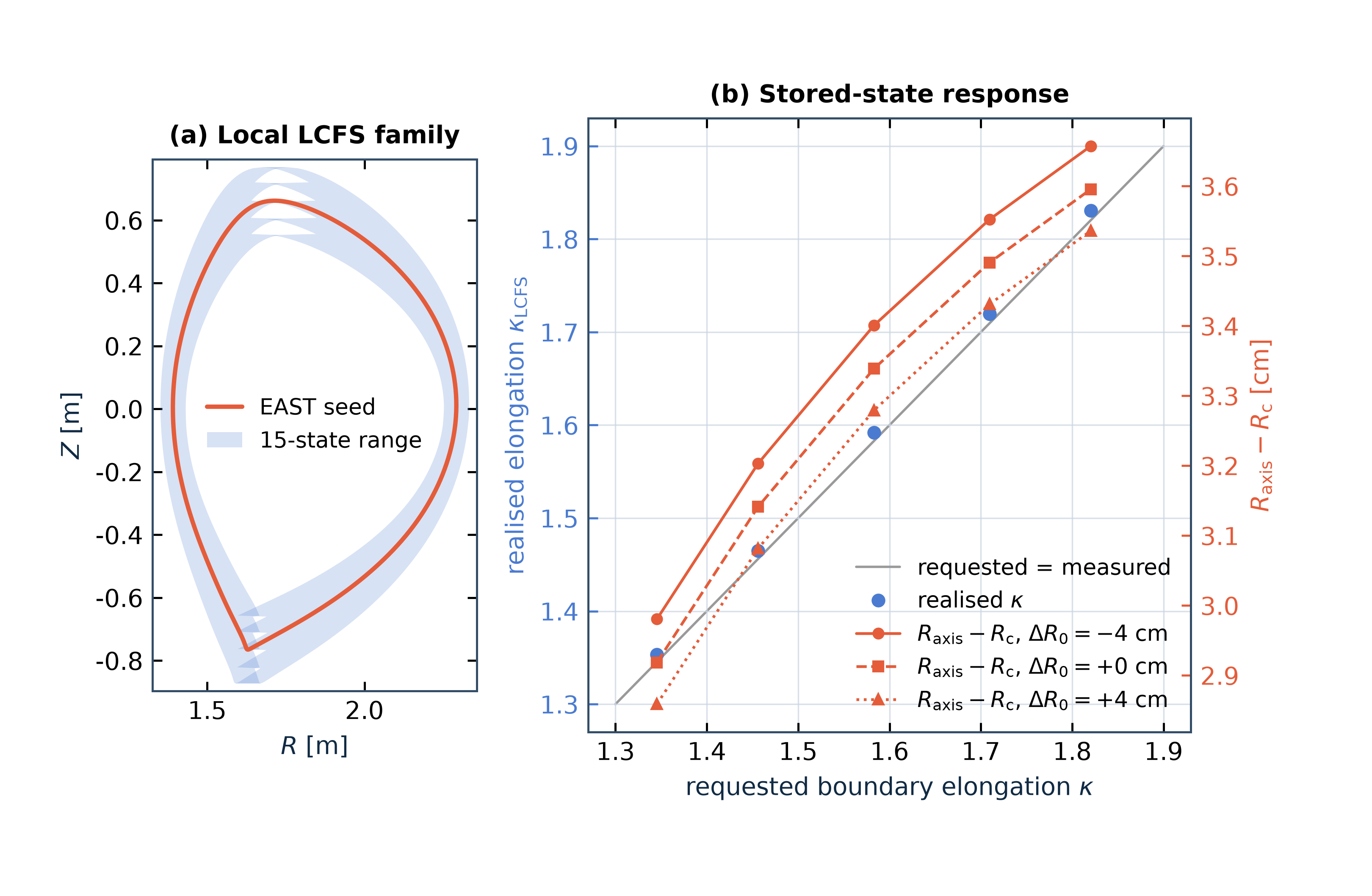}
\caption{Local fixed-boundary family grown from the EAST reference baseline.
(a) The red curve is the seed LCFS; the translucent blue band is the envelope of
the inboard and outboard LCFS intersections on horizontal cuts across the 15
accepted states. Every member remains a simple closed curve. (b) Realised response of the
stored state against the requested elongation: realised elongation on the left axis
with the identity line for reference, and the separation between the solved
magnetic axis and the prescribed boundary centre on the right axis, one series per
boundary-centre displacement. All 15 geometric displacement combinations were
accepted, so this is the sampled set and not an acceptance boundary. The three
common source-scaling levels produce duplicate states and are not counted as
independent equilibria. The X-point location is not a constraint of this solve and
is not asserted for any member.}
\label{fig:ch3-family}
\end{figure*}

Figure~\ref{fig:ch3-family} shows such a family grown from the EAST baseline of
Sec.~\ref{sec:ch3-gfile}. Elongation and boundary-centre position are displaced on
a $5\times3$ grid: the elongation over $\pm15\%$ and the boundary centre over
$\pm4$~cm. All 15 distinct states were accepted, with a maximum projected
residual of $1.6\times10^{-10}$, nested interior flux surfaces, and a positive
sampled Jacobian in every case. The original enumeration also applied three
common scale factors to both $P_\psi$ and $FF_\psi$ at fixed $I_p$. That shared
factor is absorbed by the source normalization, so the resulting 45 records
collapse to the same 15 equilibria within numerical precision and are counted
once here.

Two properties of the family are worth separating. The measured elongation tracks
the requested value closely, from $\kappa=1.35$ to $1.83$ against a requested
$1.345$ to $1.820$, while the measured minor radius stays fixed at $0.448$~m and
the on-axis pressure rises from $17.4$ to $23.4$~kPa across the geometric scan. The
pressure variation is associated with elongation and boundary position, not with
the redundant common source scaling. The separation between the solved magnetic
axis and the prescribed boundary centre stays between $2.9$ and $3.7$~cm and
responds mainly to elongation: across the full $8$~cm boundary-centre translation
at fixed elongation, $R_{\mathrm{axis}}-R_{\mathrm c}$ changes by only $0.12$~cm.
That is the expected behaviour of a fixed-boundary formulation, where translating
the prescribed surface is not the same operation as changing the equilibrium's
Shafranov shift, and it is why the two quantities are reported separately.
The acceptance envelope is wider than this grid, so Fig.~\ref{fig:ch3-family}
reports the accepted set and its response rather than a boundary of validity;
locating that boundary requires a separate, wider study.

The family size is an outcome of the same acceptance test applied to every record:
perturbations that self-intersect, invert the Jacobian or destroy flux-surface
nesting are rejected. A displaced member is not assumed to preserve the source
X-point topology, which is not constrained by the fixed-boundary solve.

\subsection{Numerical quality and storage footprint}\label{sec:ch3-quality}

Across the release, the median projected residual is $3.68\times10^{-11}$ and the
maximum is $1.15\times10^{-7}$. Acceptance also requires finite physical
quantities, positive $F^2$, a positive sampled geometric Jacobian, nested flux
surfaces, and selected safety-factor bounds. These tests establish numerical
admissibility within the chosen representation; they do not establish MHD
stability or agreement with an experimental reconstruction.

The 13,291 JSON records occupy 41~MB before archive compression and 18~MB in
the distributed archives, or 1.4~kB per equilibrium.

All 13,291 records were loaded from their stored coefficients, re-encoded,
and re-evaluated on a $96\times128$ grid: every record reproduced its stored values
under a mixed tolerance of $10^{-12}$~m absolute and $10^{-9}$ relative, every one
kept a strictly increasing normalized flux, and every one returned a positive
Jacobian away from the axis and finite pressure, safety factor and field profiles.
No record failed. This verifies that the stored coefficients are sufficient to
rebuild the equilibrium they encode; it does not re-solve the Grad--Shafranov
equation, and it is therefore not a second convergence test.

\section{Discussion and conclusions}\label{sec:discussion}

VEQDB places direct fixed-boundary solutions and G-EQDSK-derived equilibria in a
common reconstructible state while retaining their provenance. A projected
residual measures convergence of the VEQ solution, whereas a flux-map difference
measures fidelity to an input reconstruction; neither substitutes for the other.
The cross-device tests show that one coefficient representation accommodates
spherical, conventional and reactor-scale equilibria.

The database supports representation studies, numerical benchmarks and
data-driven workflows requiring a common fixed-boundary state. Parameter-sampled
records extend geometric and profile coverage, but do not represent an empirical
distribution of tokamak operation. Reference-derived model equilibria likewise
remain distinct from experimental EFIT reconstructions.

Several limitations follow from the chosen scope. The records describe the plasma
interior and LCFS, not the exterior field or open-field-line region. Conversion
accuracy is bounded by the resolution and internal consistency of the source file.
For the present G-EQDSK examples, refinement of the LCFS harmonic order reaches a
visible and current-error plateau. The likely causes include the non-orthogonal,
nonlinear MXH parameterization and the balance imposed between the flux and LCFS
residual blocks; a broader study is required to separate these effects. Local
perturbations preserve an accepted fixed-boundary geometry but do not constrain
the X-point position. Finally, the archive reload test verifies serialization and
reconstruction, not force balance or MHD stability. Future releases should expand
verified experiment-derived coverage and add exterior-field information while
retaining explicit provenance and route-specific validation.

VEQDB provides 13,291 accepted fixed-boundary equilibria across 267 devices in a
common MXH--Chebyshev representation. The three record classes share a schema
while retaining distinct provenance; cross-device comparisons quantify source
fidelity and archive-wide reload tests verify reconstruction from stored roots.
For the representative $129\times129$ G-EQDSK inputs, compact JSON reduces the
per-equilibrium file size by factors of 89--96 while retaining reproducible
evaluation on application-selected grids. The 18~MB archive, schema and
supporting software are available at \url{https://github.com/FusionAlpha/veqdb}.
\appendix
\section{Compact JSON field reference}\label{app:json-fields}

The public compact-JSON writer emits one flat object in the field order given in Table~\ref{tab:json-fields}. With the default \code{header="Equilibrium"} and no active \code{K_max}, an object has 31 fields: 19 reconstruction fields, 11 display-only diagnostics, and one description. An active \code{K_max} is inserted immediately after \code{Nt}, increasing the count to 32. Passing \code{header=None} omits only the description. The reader extracts the reconstruction fields and ignores the description, all display diagnostics, and any other extra fields. Missing diagnostics therefore do not prevent reconstruction, whereas every applicable reconstruction field remains required.

The default \code{precision=10} writes at most ten significant digits. Passing \code{precision=None} preserves the source float64 values in decimal form. Both modes require finite floating-point values and preserve signed zero. Exact trailing positive zeros are omitted from coefficient vectors. The cosine and sine coefficient families additionally omit trailing all-positive-zero harmonic rows; an all-zero row between retained harmonics is written as \code{[0.0]} so that its index is not lost. On input, the reader constructs float64 arrays and restores omitted trailing entries by zero-padding to the common radial coefficient width. The \code{c_lcfs} and \code{c_coeffs} arrays use cosine orders $c_0,c_1,\ldots$, whereas \code{s_lcfs} and \code{s_coeffs} use sine orders $s_1,s_2,\ldots$.

\footnotesize
\begin{longtable}{@{}>{\raggedright\arraybackslash}p{0.18\textwidth}>{\raggedright\arraybackslash}p{0.17\textwidth}>{\raggedright\arraybackslash}p{0.57\textwidth}@{}}
  \caption{Compact JSON fields in their exact output order. The conditional \code{K_max} row is present only when its regularity cap changes a represented harmonic.}\label{tab:json-fields}                \\
  \toprule
  Field               & Role                      & Definition                                                                                                                                              \\
  \midrule
  \endfirsthead
  \caption[]{Compact JSON fields (continued).}                                                                                                                                                              \\
  \toprule
  Field               & Role                      & Definition                                                                                                                                              \\
  \midrule
  \endhead
  \midrule
  \multicolumn{3}{r}{Continued on next page}                                                                                                                                                                \\
  \endfoot
  \bottomrule
  \endlastfoot
  \code{header}       & Description               & Optional human-readable string; the public default is \code{Equilibrium}. It is not a reconstruction input.                                             \\
  \code{R_axis}       & Diagnostic                & Magnetic-axis major-radius coordinate, in m.                                                                                                            \\
  \code{Z_axis}       & Diagnostic                & Magnetic-axis vertical coordinate, in m.                                                                                                                \\
  \code{B_axis}       & Diagnostic                & Signed toroidal magnetic field at the magnetic axis, in T.                                                                                              \\
  \code{P_axis}       & Diagnostic                & Pressure at the magnetic axis, in Pa.                                                                                                                   \\
  \code{B0}           & Reconstruction            & Signed reference toroidal field, in T; it anchors the LCFS field function through $F(1)=R_0B_0$.                                                        \\
  \code{P0}           & Reconstruction            & LCFS pressure $P(1)$, in Pa.                                                                                                                            \\
  \code{R0}           & Reconstruction            & Major-radius coordinate of the boundary reference frame, in m.                                                                                          \\
  \code{Z0}           & Reconstruction            & Vertical coordinate of the boundary reference frame, in m.                                                                                              \\
  \code{a}            & Reconstruction            & Minor-radius scale of the boundary reference frame, in m.                                                                                               \\
  \code{betat}        & Diagnostic                & Dimensionless toroidal beta, $2\mu_0\langle P\rangle/B_0^2$.                                                                                            \\
  \code{Ip}           & Diagnostic                & Total plasma current, in A. The G-EQDSK comparison records use the explicit positive-current convention.                                               \\
  \code{qmin}         & Diagnostic                & Signed minimum of the safety-factor values on the stored radial grid.                                                                                   \\
  \code{q0}           & Diagnostic                & Signed safety factor at the magnetic-axis node, $q[0]$.                                                                                                 \\
  \code{q95}          & Diagnostic                & Safety factor linearly interpolated at normalized poloidal flux $\psi_{\mathrm n}=0.95$.                                                                \\
  \code{Wth}          & Diagnostic                & Thermal energy $\tfrac{3}{2}\int P\,\dd V$, in J.                                                                                                       \\
  \code{psi_lcfs}     & Diagnostic                & LCFS poloidal flux $\psi[-1]$ in the axis-zero gauge, in Wb/rad. The comparison records use an outward-increasing flux convention.                    \\
  \code{kappa_lcfs}   & Reconstruction            & Dimensionless LCFS elongation $\kappa(1)$.                                                                                                              \\
  \code{c_lcfs}       & Reconstruction            & Dimensionless LCFS cosine phase coefficients $[c_0,c_1,\ldots]$.                                                                                        \\
  \code{s_lcfs}       & Reconstruction            & Dimensionless LCFS sine phase coefficients $[s_1,s_2,\ldots]$.                                                                                          \\
  \code{Nr}           & Configuration             & Number of radial Gauss--Lobatto nodes; an integer of at least four.                                                                                     \\
  \code{Nt}           & Configuration             & Number of uniformly spaced periodic poloidal nodes; an integer of at least four and sufficient to resolve the retained harmonics.                       \\
  \code{K_max}        & Conditional configuration & Optional cap in $K_m=\min(m,K_{\max})$. It is omitted when unset or when $K_{\max}\geq M_{\max}$, because the cap then changes no represented harmonic. \\
  \code{h_coeffs}     & Reconstruction            & One-dimensional Chebyshev coefficients for the interior horizontal-shift profile $h(r)$.                                                                \\
  \code{v_coeffs}     & Reconstruction            & One-dimensional Chebyshev coefficients for the interior vertical-shift profile $v(r)$.                                                                  \\
  \code{kappa_coeffs} & Reconstruction            & One-dimensional Chebyshev coefficients for the interior correction to $\kappa(r)$ about \code{kappa_lcfs}.                                              \\
  \code{c_coeffs}     & Reconstruction            & Array of Chebyshev-coefficient rows for the interior cosine phase profiles, ordered $c_0,c_1,\ldots$.                                                   \\
  \code{s_coeffs}     & Reconstruction            & Array of Chebyshev-coefficient rows for the interior sine phase profiles, ordered $s_1,s_2,\ldots$.                                                     \\
  \code{P_psi}        & Reconstruction            & Samples of $P_\psi=\dd P/\dd\psi$ on the \code{Nr} radial nodes, in A/m$^3$.                                                                            \\
  \code{FF_psi}       & Reconstruction            & Samples of $FF_\psi=\dd(F^2/2)/\dd\psi$ on the \code{Nr} radial nodes, in T.                                                                            \\
  \code{psi_r}        & Reconstruction            & Samples of $\psi_r=\dd\psi/\dd r$ on the \code{Nr} radial nodes, in Wb/rad.                                                                             \\
  \code{psi_rr}       & Reconstruction            & Independent samples of $\psi_{rr}=\dd^2\psi/\dd r^2$ on the \code{Nr} radial nodes, in Wb/rad.                                                          \\
\end{longtable}
\normalsize

Listing~\ref{lst:iter-json} shows the ITER inverse-solve record used in the
cross-device comparison. Scalar values use the writer's default ten-significant-
digit precision; each array retains its first stored value and uses an ellipsis
for the omitted suffix. The ellipses are explanatory notation and are not valid
JSON tokens, so the listing documents the interface but is not itself a
deserializable record.
By default, deserialization restores \code{Nr} and \code{Nt} from the object.
Supplying either value explicitly requests another Lobatto/spectral
materialization; changing \code{Nr} linearly interpolates all four physical root
profiles in the geometric radius $r$, while the shape coefficients are unchanged.

\begin{lstlisting}[
  basicstyle=\ttfamily\footnotesize,
  columns=fullflexible,
  keepspaces=true,
  frame=single,
  caption={Abbreviated compact JSON record for the ITER inverse-solve equilibrium.},
  label={lst:iter-json}
]
{
  "header": "ITER-high",
  "R_axis": 6.41433755,
  "Z_axis": 0.567475304,
  "B_axis": 5.287156252,
  "P_axis": 1082363.656,
  "B0": 5.343738403,
  "P0": 0,
  "R0": 6.149253112,
  "Z0": 0.1558981237,
  "a": 2.078375134,
  "betat": 0.02650555021,
  "Ip": 15367207.62,
  "qmin": 0.9752347351,
  "q0": 0.9752347351,
  "q95": 2.850653255,
  "Wth": 369443343.6,
  "psi_lcfs": 13.1497591,
  "kappa_lcfs": 1.704561888,
  "c_lcfs": [-0.0536336596, ...],
  "s_lcfs": [0.2527035586, ...],
  "Nr": 32,
  "Nt": 32,
  "h_coeffs": [0.1403912877, ...],
  "v_coeffs": [0.5589676558, ...],
  "kappa_coeffs": [-0.7074466849, ...],
  "c_coeffs": [[0.03967352887, ...], ...],
  "s_coeffs": [[-0.2567886147, ...], ...],
  "P_psi": [-136475.83, ...],
  "FF_psi": [-4.43542401, ...],
  "psi_r": [0.0, ...],
  "psi_rr": [33.4511177, ...]
}
\end{lstlisting}

\bibliographystyle{elsarticle-num}
\bibliography{references}
\end{document}